\documentstyle[epsf,preprint,tighten,prd,aps]{revtex}

\begin{document}
%\epsfverbosetrue
\draft

%%% Front matter

\title{
  Chaotic Scattering and Capture of Strings by Black Hole
}
\author{
  Andrei V. Frolov\footnotemark%
    \thanks{Electronic address: \texttt{andrei@phys.ualberta.ca}}
  and Arne L. Larsen\footnotemark%
    \thanks{Electronic address: \texttt{all@fysik.ou.dk}}
}
\address{
  \medskip
  \footnotemark[1]
  Physics Department, University of Alberta\\
  Edmonton, Alberta, Canada, T6G 2J1\\
  \smallskip
  \footnotemark[3]
  Physics Department, University of Odense\\
  Campusvej 55, 5230 Odense M, Denmark
}
\date{August 11, 1999}
\maketitle

\begin{abstract}
  We consider scattering and capture of circular cosmic strings by a
  Schwarzschild black hole. Although being a priori a very simple
  axially symmetric two-body problem, it shows all the features of
  chaotic scattering. In particular, it contains a fractal set of
  unstable periodic solutions; a so-called strange repellor. We study
  the different types of trajectories and obtain the fractal dimension
  of the basin-boundary separating the space of initial conditions
  according to the different asymptotic outcomes. We also consider the
  fractal dimension as a function of energy, and discuss the transition
  from order to chaos.
\end{abstract}


%%% Main body

Chaos in general relativity and cosmology is by now a well-established
subject. Some studies, pioneered by the work of Hawking \cite{haw} and
Page \cite{page}, have been concerned with the behavior of solutions to
the Einstein equations themselves; the most famous example now being
the mixmaster universe (see for instance \cite{neil1}, and references
given therein). In other studies, the subject of interest has been the
behavior of test-particle trajectories in black hole spacetimes (see
for instance \cite{dettmann,gibbons}). In the latter case, since
point-particle motion is completely integrable in the generic
Kerr-Newman background \cite{carter}, it was necessary to consider
quite complicated multi black hole spacetimes (typically of the
Majumdar-Papapetrou type \cite{maj,pap}) to obtain chaotic
point-particle dynamics. Taking into account that these systems 
represent (at least) three-body problems, and comparing with Newtonian
dynamics, it was certainly no surprise that chaotic dynamics appeared.
(It should be mentioned, however, that in the two black hole case,
chaos is really a relativistic effect \cite{dettmann}.) The main
interest in these systems therefore was also to use and further develop
coordinate-invariant descriptions and measures of chaotic behavior
\cite{ott,neil2}, suitable for general relativity where space and time
are not absolute.

\begin{figure}
  \centerline{\epsfysize=3.8cm \epsfbox{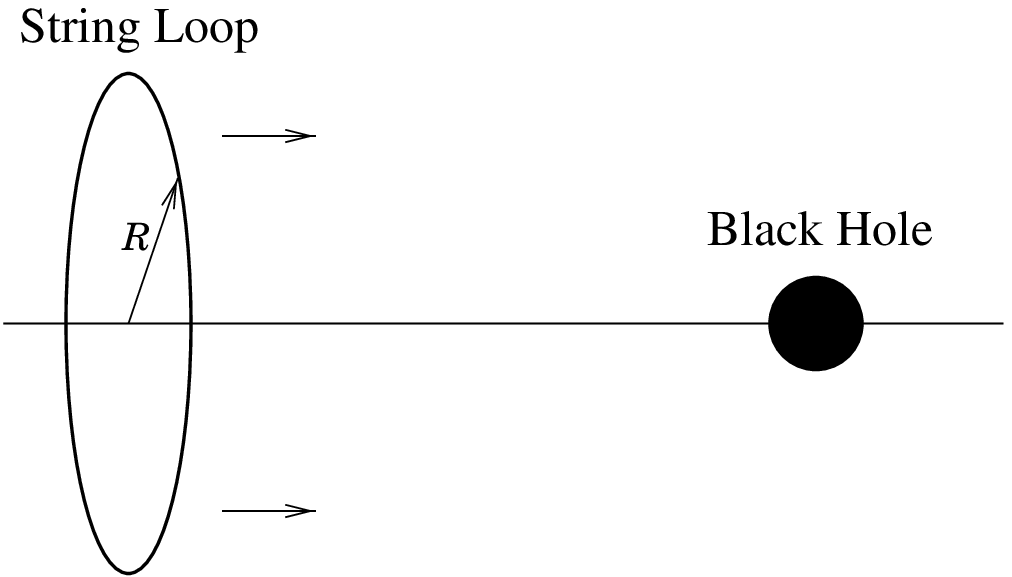}}
  \caption{
    Schematic illustration of string loop approaching black hole.
  }
  \label{fig:string}
\end{figure}

In the present paper, we study a conceptually simpler and more
symmetric system in general relativity that nevertheless, and maybe
somewhat surprisingly, leads to chaotic behavior. We consider the
two-body problem of a circular test-string in the background of a
Schwarzschild black hole. The circular string is taken to be coaxial
with the black hole and is allowed to oscillate in its plane and to
propagate in the direction perpendicular to its plane, as illustrated
in Fig.~\ref{fig:string}. The combined system is therefore axially
symmetric.

The physical picture we have in mind is that of a cosmic string nearby
an astrophysical black hole. For a GUT string with string tension of
the order $G\mu\approx 10^{-6}$ \cite{vilenkin}, the test-string
approximation should be valid even for a string initially up to (say)
$10^4$ times longer than the black hole horizon radius. Moreover, for a
GUT string we can use the leading order thickness approximation, where
the cosmic string is described by the Nambu-Goto action
\cite{vilenkin}
\begin{equation} \label{eq:action}
  S = -\mu \int d\tau\, d\sigma\, \sqrt{-\det[G_{ab}]}\;,
\end{equation}
where $G_{ab}=g_{\mu\nu}X^\mu_{,a}X^\nu_{,b}$ is the induced metric on
the string worldsheet. In this case, the string equations of motion
and constraints (in conformal gauge) take the standard form
\begin{eqnarray}
  & \ddot{X}^\mu-X''^\mu + \Gamma^\mu_{\rho\sigma}
    (\dot{X}^\rho \dot{X}^\sigma - X'^\rho X'^\sigma) = 0,& \label{eq:eom}\\
  & g_{\mu\nu} \dot{X}^\mu X'^\nu =
    g_{\mu\nu} (\dot{X}^\mu \dot{X}^\nu + X'^\mu X'^\nu) = 0,& \label{eq:constraint}
\end{eqnarray}
where dot $\dot{\ }$ and prime $\ '$ denote derivatives with respect to
string worldsheet coordinates $\tau$ and $\sigma$. Using Schwarzschild coordinates for the background
metric
\begin{equation} \label{eq:bg}
  ds^2 =
    -\left(1 - \frac{2M}{r}\right) dt^2
    + \left(1 - \frac{2M}{r}\right)^{-1} dr^2
    + r^2\, d\Omega^2,
\end{equation}
and parameterizing the circular string by ansatz
\begin{equation} \label{eq:ansatz}
  t=t(\tau), \hspace{1em}
  r=r(\tau), \hspace{1em}
  \theta=\theta(\tau), \hspace{1em}
  \phi=\sigma,
\end{equation}
one finds that the string equations of motion (\ref{eq:eom}) and
constraints (\ref{eq:constraint}) lead to the following system of
ordinary differential equations
\begin{equation} \label{eq:t}
  \dot{t} = E \left(1 - \frac{2M}{r}\right)^{-1},
\end{equation}
\begin{equation} \label{eq:r}
  \ddot{r} = (r-3M) {\dot{\theta}}^2 - (r-M) \sin^2\theta,
\end{equation}
\begin{equation} \label{eq:theta}
  \ddot{\theta} = -\frac{2}{r}\, \dot{r} \dot{\theta} - \sin\theta \cos\theta,
\end{equation}
supplemented by the constraint
\begin{equation} \label{eq:E}
  \dot{r}^2 + (r^2-2Mr) ({\dot{\theta}}^2 + \sin^2\theta) = E^2.
\end{equation}
The integration constant $E$, appearing here, will play the role of
external control parameter (``order parameter'') for the system. It is
equal to the total conserved energy of the string divided by $2\pi\mu$.
In the absence of first integrals other than (\ref{eq:E}), we are thus
dealing with a three-dimensional phase space. Notice also that the
worldsheet time $\tau$ is not equal to the proper time. The worldsheet
time $\tau$ is however finite for a string falling into the black hole.

Eqs. (\ref{eq:t}--\ref{eq:E}) are equivalent to the following
Hamiltonian system
\begin{equation} \label{eq:H}
  H = \frac{1}{2} \left(1-\frac{2M}{r}\right) P_{r}^2
      + \frac{1}{2r^2}\, P_{\theta}^2
      - \frac{E^2}{2} \left(1-\frac{2M}{r}\right)^{-1}
      + \frac{1}{2}\, r^2 \sin^2\theta,
\end{equation}
with the constraint $H=0$. This is very similar to the Hamiltonian
describing zero angular momentum photons in the Schwarzschild
background \cite{mtw}, except for the last term in the potential which
is due to the string tension. This is precisely the term that leads to
non-integrability and, as we shall see, chaos.

The Hamiltonian system (\ref{eq:H}) has been previously considered and
solved in the equatorial plane ($\theta=\pi/2$) \cite{larsen}, and some
trajectories in the general case were computed numerically in
\cite{vega}. In this paper, we make a more complete analysis of the
dynamics associated with the Hamiltonian (\ref{eq:H}). In particular,
we shall demonstrate the presence of chaos in this simple system.

Due to the non-integrable nature of the dynamical system
(\ref{eq:t}--\ref{eq:E}), the analysis of the string evolution was done
numerically. The string trajectories for various initial conditions
were obtained by integrating Eqs. (\ref{eq:r}) and (\ref{eq:theta})
using the fifth order embedded Runge-Kutta method with adaptive step
size control \cite{nr}. The constraint (\ref{eq:E}) was used to
independently check for numerical precision.

\begin{figure}
  \begin{center}
    \begin{tabular}{c@{\hspace{3em}}c}
      \epsfysize=3.8cm \epsfbox{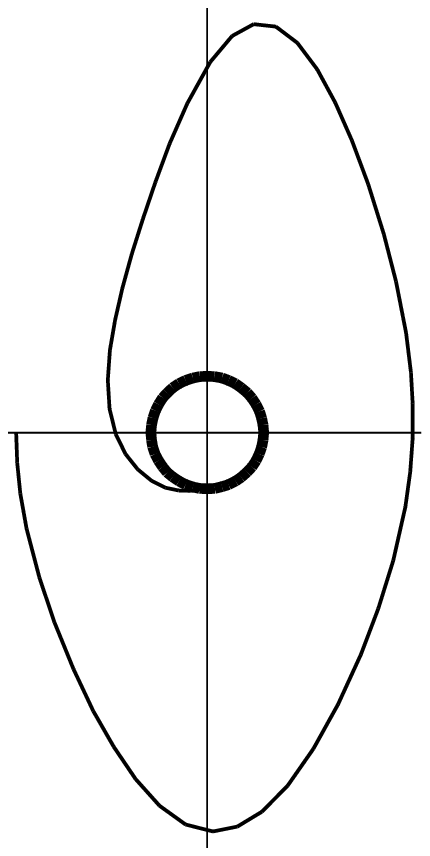} &
      \epsfysize=3.8cm \epsfbox{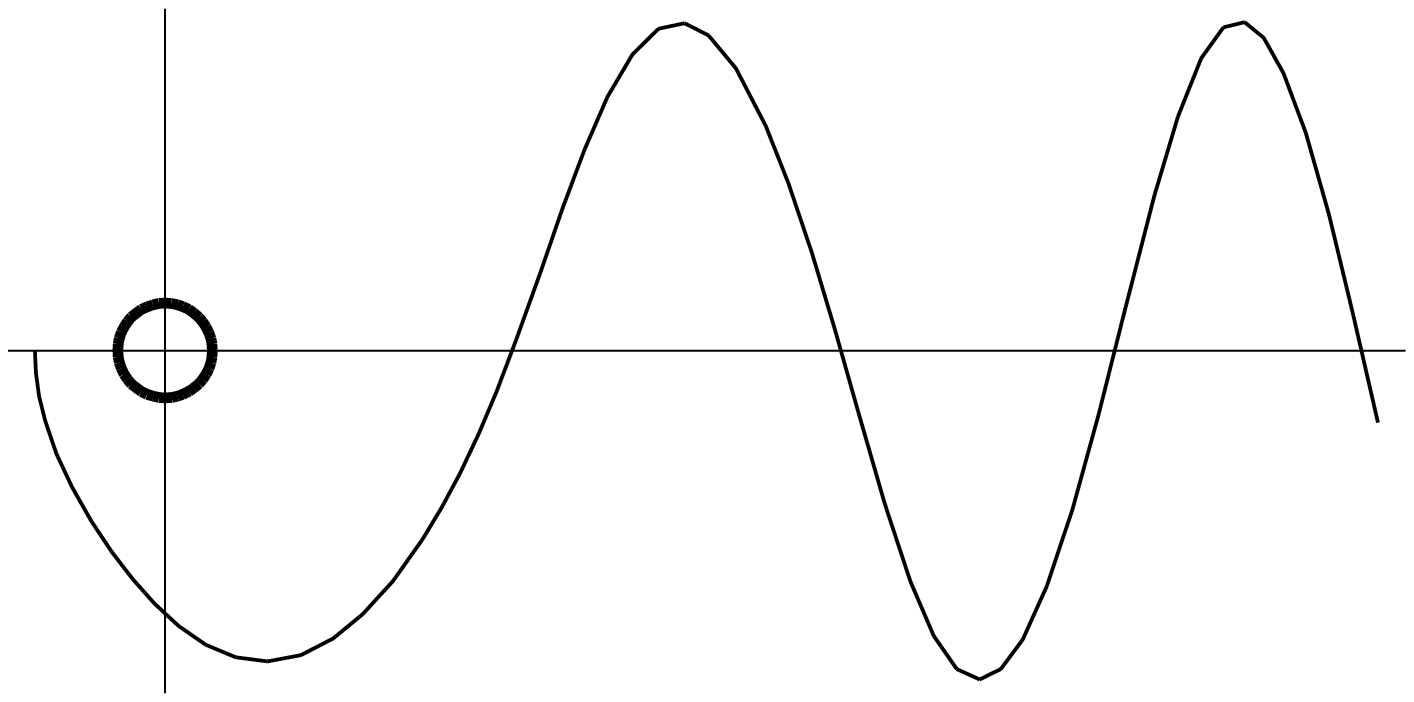} \\
      (a) & (b) \\ \\
    \end{tabular}
    \epsfysize=3.8cm \epsfbox{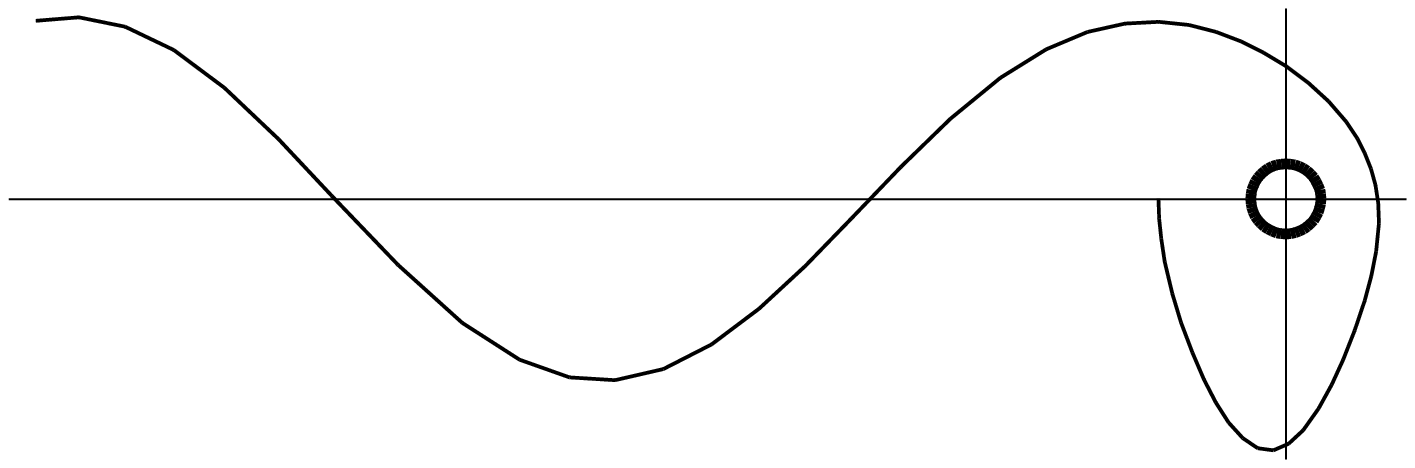} \\ (c) \\
  \end{center}
  \caption{
    Possible asymptotic outcomes of the string evolution:
    (a) capture, (b) escape, and (c) escape with backscatter.
    Axes are $(r \cos\theta, r \sin\theta)$, and the thick
    circle represents the event horizon of the black hole.
  }
  \label{fig:outcomes}
\end{figure}

It is easy to see that there are three possible asymptotic outcomes of
the string dynamics. The string can either fly by the black hole and
escape to $(r,\theta)=(\infty, \pi)$, or the string can backscatter and
escape to $(r,\theta)=(\infty, 0)$, or the string can be captured by
the black hole $r\leq 2M$. Some examples of string trajectories
illustrating these outcomes are shown in Fig.~\ref{fig:outcomes}, where
we plot the string radius $R=r\sin\theta$ (vertically) as a function of
$Z=r\cos\theta$ (horizontally, string comes in from the left). In the
three examples shown, the string is initially collapsed at a position a
few horizon radii outside of the black hole. It then expands and
propagates to the right --- towards the black hole. In
Fig.~\ref{fig:outcomes}a, the string passes the black hole but then
returns and is captured, $r\leq 2M$. In Fig.~\ref{fig:outcomes}b, the
string passes the black hole and escapes to the right,
$(r,\theta)=(\infty,\pi)$. In Fig.~\ref{fig:outcomes}c, the string
passes the black hole, then returns and finally escapes to the left, 
$(r,\theta)=(\infty,0)$.

\begin{figure}
  \begin{center}
    \begin{tabular}{c@{\hspace{1em}}c@{\hspace{1em}}c@{\hspace{1em}}c@{\hspace{1em}}c}
      \epsfysize=3.8cm \epsfbox{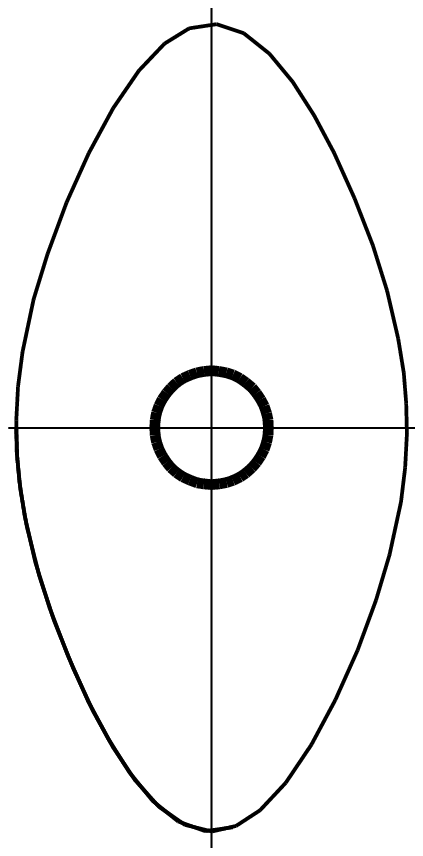} &
      \epsfysize=3.8cm \epsfbox{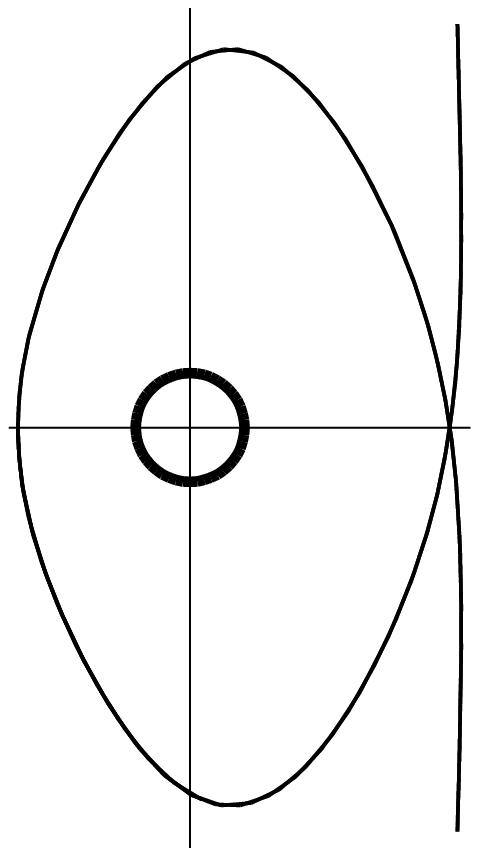} &
      \epsfysize=3.8cm \epsfbox{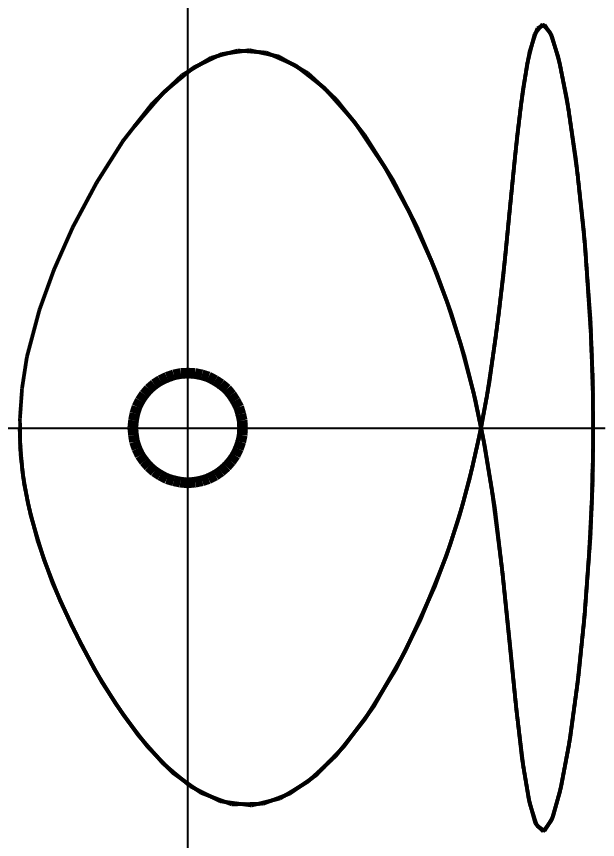} &
      \epsfysize=3.8cm \epsfbox{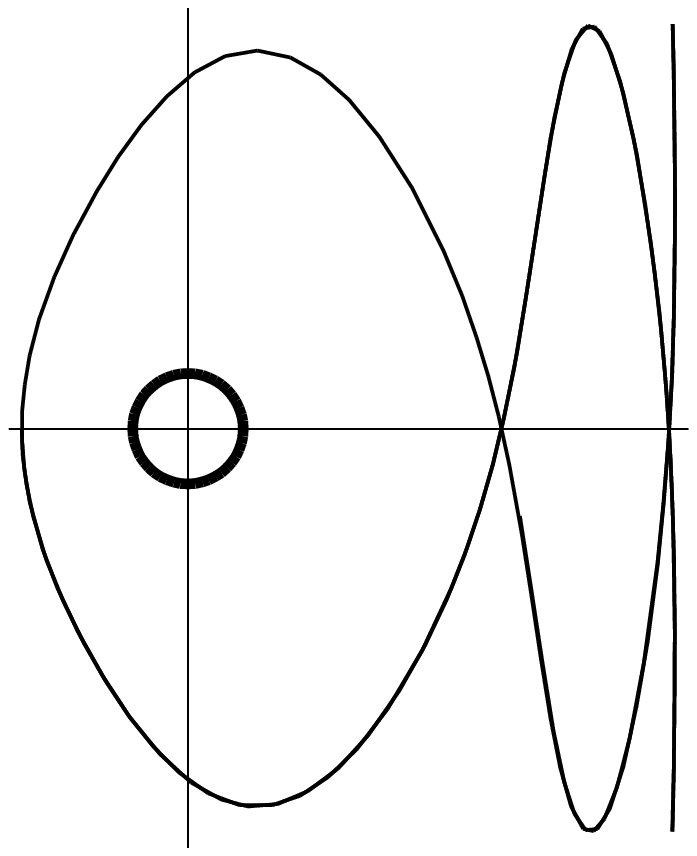} &
      \epsfysize=3.8cm \epsfbox{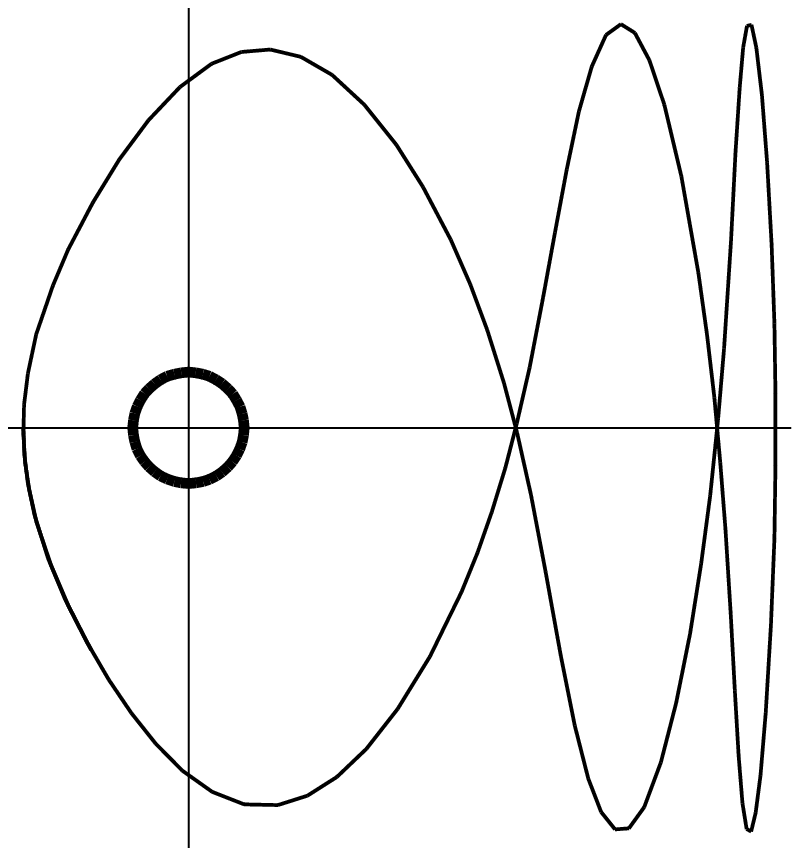} \\
      (a) & (b) & (c) & (d) & (e) \\
    \end{tabular}
  \end{center}
  \caption{
    Unstable periodic orbits.
  }
  \label{fig:period}
\end{figure}

Besides the solutions with these three asymptotic outcomes, there is an
infinite set of unstable periodic orbits, which separate the solutions
with different asymptotic outcomes in the phase space of all solutions.
Typical examples of periodic orbits are shown in Fig.~\ref{fig:period}.
Besides the ones shown, there are also periodic orbits when the string
starts a little further away from the black hole and oscillates a
number of times before reaching it. There are no stable periodic orbits
in our system.

To get a better understanding of the string dynamics, we now consider a
two-dimensional slice of the four-dimensional space of initial
conditions in more detail. It is most convenient to fix the constant
``energy'' $E$ and then to impose one more relation between initial
values of $(r,\theta,\dot{r},\dot{\theta})$ at $\tau=0$. Following the
procedure of the basin-boundary method for chaotic scattering
\cite{ott}, we then color this two-dimensional slice of initial
conditions according to the three different asymptotic outcomes
mentioned above. Due to numerical reasons, the determination of the two
asymptotic outcomes corresponding to escape is done not at infinity but
at some large, but finite, value of $r$. While in general the string
can pass this cutoff radius in one direction but change its mind later,
for sufficiently large $r$, this procedure will only lead to a wrong
color for very few points, as follows from the asymptotic behavior of
the potential in the Hamiltonian (\ref{eq:H}) (this should be
contrasted with the case of the mixmaster universe \cite{levin}, where
all the trajectories eventually bounce back).

\begin{figure}
  \begin{center}
    \begin{tabular}{c@{\hspace{2em}}c}
      \epsfbox{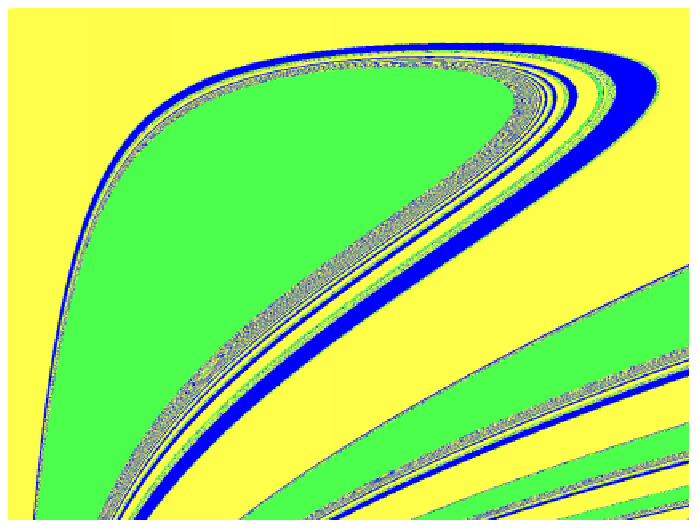} & \epsfbox{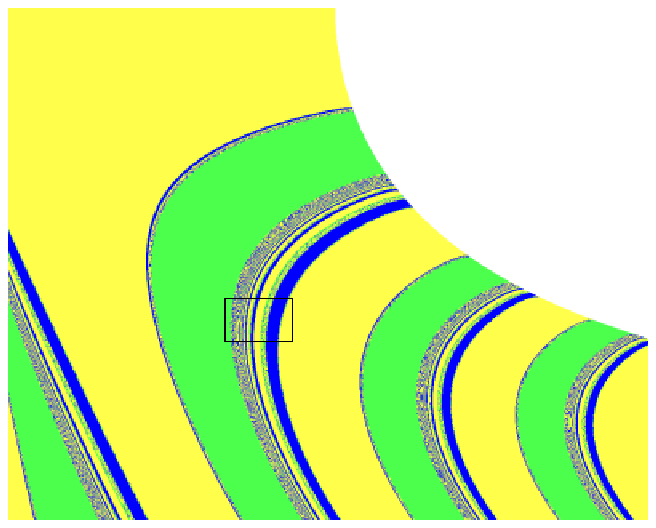} \\
      (a) & (b) \\
      \\
      \epsfbox{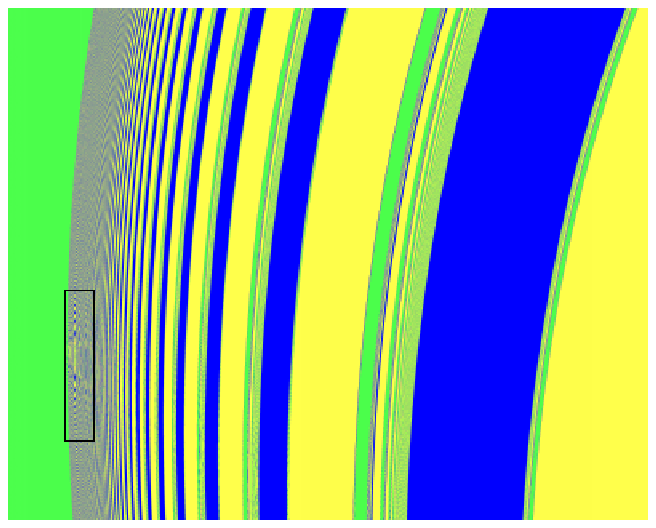} & \epsfbox{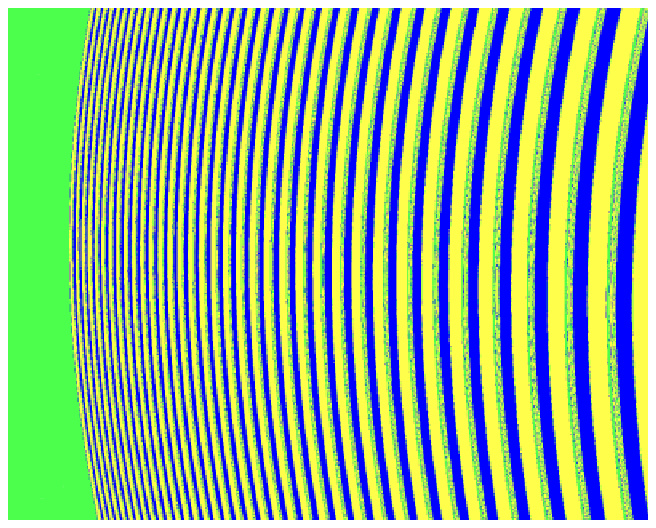} \\
      (c) & (d) \\
      \\
    \end{tabular}
    \\
    {\bf Legend}:
      \epsfbox{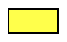}\,\ capture,
      \epsfbox{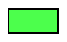}\,\ escape,
      \epsfbox{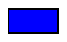}\,\ escape with backscatter.
  \end{center}
  \caption{
    Basin-boundaries, plotted for $E = 14.0 M$:
    (a) slice $\theta=0$,
        with $r \in [2.0 M, 29.2 M]$ on horizontal axis,
        and $-\dot{r} \in [0, E]$ on vertical axis;
    (b) slice $\frac{d\ }{d\tau}(r \cos\theta) = 0$,
        with $r \in [2.0 M, 27.5 M]$ on horizontal axis,
	and $\theta \in [0, \pi/2]$ on vertical axis;
    (c) and (d) show fractal detail of figures (b) and (c) respectively.
  }
  \label{fig:phase}
\end{figure}

In Fig.~\ref{fig:phase} we show examples of this coloring procedure
applied to the phase space of string solutions. Fig.~\ref{fig:phase}a
and Fig.~\ref{fig:phase}b show two different two-dimensional slices of
the space of initial conditions. Fig.~\ref{fig:phase}c and
Fig.~\ref{fig:phase}d show magnifications of the regions of
Fig.~\ref{fig:phase}b and Fig.~\ref{fig:phase}c indicated in the
previous figures. The boundaries between different colors in these
figures correspond to the unstable periodic orbits. The magnifications
of these so-called basin-boundaries reveal a fractal structure, see
Figs.~\ref{fig:phase}b-d. This provides the coordinate-invariant
indication that the dynamics is in fact chaotic \cite{ott}. 

\begin{figure}
  \centerline{\epsfysize=7.5cm \epsfbox{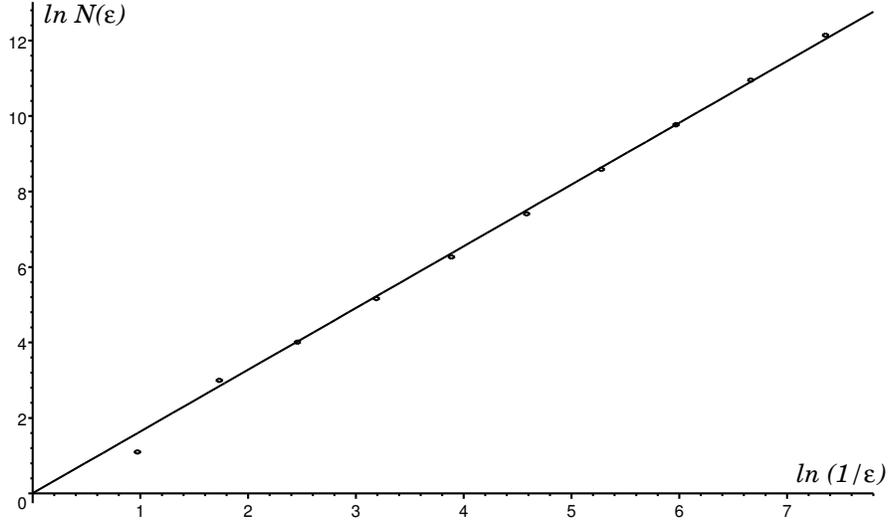}}
  \caption{
    Number of fractal points as a function of resolution,
    calculated for slice in Fig.~\ref{fig:phase}b.
  }
  \label{fig:scale}
\end{figure}

To get a quantitative measure we can also determine the fractal
dimension. Consider for instance Fig.~\ref{fig:phase}b, which
corresponds to the two-dimensional slice in the space of initial
conditions, given by
\begin{equation} \label{eq:slice}
  E = 14.0 M, \hspace{1em}
  \frac{d\ }{d\tau}(r \cos\theta) = 0
  \text{ at } \tau=0.
\end{equation}
The box-counting dimension is defined by \cite{ott}
\begin{equation} \label{eq:fdim}
  D = \lim_{\epsilon \rightarrow 0} \frac{\ln N(\epsilon)}{\ln (1/\epsilon)}\,,
\end{equation}
where $N(\epsilon)$ is the number of squares of side length $\epsilon$
needed to cover the basin-boundary. A square should be counted only if
it contains at least two different colors, otherwise it is not part of
the basin-boundary. The result of this counting is shown as a plot of
$\ln{N(\epsilon)}$ versus $\ln(1/\epsilon)$ in Fig.~\ref{fig:scale}. 
The straight line is a least-square fit to the data points and shows
that the dependence is indeed linear over the very wide range of
resolution. The fractal dimension of Fig.~\ref{fig:phase}b, calculated
from the highest-resolution grid of $4000 \times 3200$ points, is
\begin{equation}
  D = 1.65 \pm 0.03.
\end{equation}
The non-integer value shows in a coordinate-invariant way \cite{barn}
that the basin-boundary is indeed a fractal; a so-called strange
repellor. The error in the result is due to the finite size of the
numerical grid and slow convergence rate of definition (\ref{eq:fdim}),
and can be estimated by examining $D$ calculated at lower resolutions.

\begin{figure}
  \centerline{\epsfysize=7.5cm \epsfbox{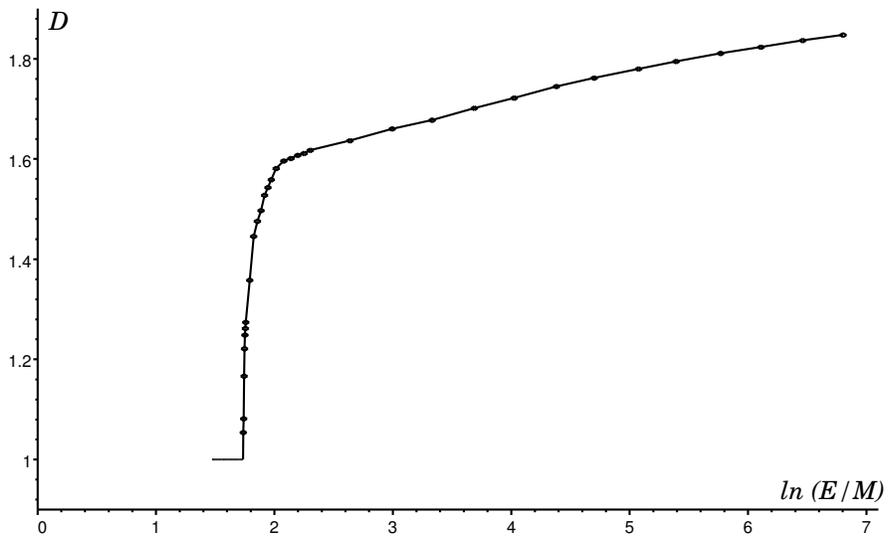}}
  \caption{
    Fractal dimension as a function of energy
    calculated for slice in Fig.~\ref{fig:phase}b.
    Radial size of the box is scaled with energy as
    $r \in [2.0 M, (2.0 + \frac{25.5}{14.0}\, \frac{E}{M}) M]$
    to keep the choice of the region consistent.
  }
  \label{fig:energy}
\end{figure}

It is interesting to also examine the fractal dimension as a function
of ``energy'' $E$. We considered the slice (\ref{eq:slice}) in the
energy range $E \in [0,10^3 M]$ and computed the fractal dimension
(\ref{eq:fdim}) of the basin-boundary (corresponding to
Fig.~\ref{fig:phase}b and its analog for different energies) as a
function of energy $E$. The result is shown in Fig.~\ref{fig:energy},
where we plot the fractal dimension $D$ as a function of $\ln(E/M)$.
For low energies, $E \lesssim 4.37M$, the dynamics is completely
regular in the sense that there is only one asymptotic outcome ---
namely capture. For slightly larger energies, with values of $E$
between $4.37M$ and $5.67M$, escape becomes possible but the different
asymptotic outcomes are connected by ``regular transitions'',
corresponding to one-dimensional curves in the space of initial
conditions (and thus corresponding to fractal dimension $1$). However,
at energy $E \simeq 5.67M$, the picture changes dramatically from
regular to highly chaotic dynamics. In a very narrow energy range, the
fractal dimension changes abruptly from $1$ to approximately $1.6$,
whereafter it increases slightly for higher energies. It must be
stressed that the numerical values given above for energies separating
the different ``phases'' are slice-dependent. The phenomenon explained
and demonstrated by Fig.~\ref{fig:energy} is however generic.

\bigskip
In conclusion, we have shown in a coordinate invariant way that the
axially symmetric system of a circular test-string in the Schwarzschild
black hole background is chaotic. Chaos sets in abruptly at a certain
``critical'' value of the external control parameter, which is related
to the conserved energy of the string. Below this critical energy, the
dynamics is regular. Above the critical energy, the dynamics is highly
irregular and chaotic.

In all fairness, it should be mentioned that we have neglected some
physical effects, including the backreaction and gravitational
radiation of the cosmic string. However, the system considered here
represents the simplest and most symmetric example of string dynamics
in black hole spacetimes and therefore suggests quite generally that
string dynamics in black hole spacetimes is chaotic (notice however
that the problem of {\it stationary} strings in black hole spacetimes
is completely integrable \cite{zel}). It would thus be interesting to
consider other dynamical string configurations as well, and we notice
that there has recently been some interest in the scattering and
capture of open strings by a black hole; see Refs.~\cite{fro,page2} and
references given therein.

\section*{Acknowledgements}
We would like to thank V. P. Frolov and D. N. Page for stimulating
discussions. This research was supported by the Natural Sciences and
Engineering Research Council of Canada and by the Killam Trust.

%%% Bibliography

%%% Figures

\end{document}